\documentclass[prl,twocolumn,showpacs,preprintnumbers,amsmath,amssymb,floatfix]
{revtex4}

\usepackage{graphicx}
\usepackage[center]{subfigure}
\usepackage{mciteplus}
\usepackage{placeins}
\mciteSetBstMidEndSepPunct{\relax}{\relax}{\relax}

\begin{document}

\title{
Modeling multiple time scales during glass formation with phase field crystals
}

\author{Joel Berry and Martin Grant}

\affiliation{McGill University, Physics Department,
3600 rue University, Montr\'eal, Qu\'ebec, Canada H3A 2T8}

\date{\today}

\begin{abstract}
The dynamics of glass formation in
monatomic and binary liquids are
studied numerically using a microscopic field theory for 
the evolution of the time-averaged atomic number density. 
A stochastic framework combining
phase field crystal free energies and
dynamic density functional theory
is shown to successfully
describe several aspects of glass formation
over multiple time scales.
Agreement with mode coupling theory is demonstrated for
underdamped liquids at moderate supercoolings,
and a rapidly growing dynamic correlation
length is found to be associated with fragile behavior.
\end{abstract}

\maketitle

A unified theoretical framework within which the glass transition may be 
understood does not currently exist.
The most significant theoretical advances have been
concentrated near the early stages of slowing, leaving the
intermediate and late stages relatively poorly understood.
Mode coupling theory \cite{gotzeMCTbook09} (MCT)
and molecular dynamics \cite{glotzer00}, for example, have
provided insight into the initial regime of slowing
above the so-called crossover temperature $T_c$,
but are ineffective when applied to the slower regimes that
occupy roughly ten orders of magnitude in time between $T_c$ and 
the glass transition temperature $T_g$.

Time- or ensemble-averaged dynamic density functional theories 
\cite{marctara99,*archerevans04,kawasaki02,*lust93,ABL06,*kimkawasaki07,
*archer09,*archer06,marconi06,*marconi07}
(DDFTs) have been
proposed as a more efficient means of describing slow dynamics 
below $T_c$, but
several key issues remain unresolved:
which of the proposed equations of motion are most appropriate,
whether the details of the free energy significantly
influence dynamics, and
whether the detailed predictions of MCT 
can be reproduced and eventually improved upon by such theories.
Mean-field DFT
functionals are known to typically produce multivalley free energy
landscapes in which an exponential number of aperiodic solid states
coexist below a certain $T$
\cite{singh85,*dasgupram92,RFOT07,pfcglass08}.
However, the nature of the transition by which a liquid
evolves toward and between these aperiodic solid states upon quenching 
is influenced heavily by the microscopic dynamics and thus, in DDFT, 
the equation of motion employed.
Approximate analytic results \cite{ABL06,*kimkawasaki07,*archer09}
indicate that two DDFT equations of motion
may describe a MCT-type glass transition, but
numerical simulations have confirmed only
stretched exponential decay and 
super-Arrhenius slowing in related, non-DDFT models 
\cite{kawasaki02,*lust93}.
Here we provide direct numerical
solutions for a candidate DDFT that considers both inertia and damping,
and utilizes the simplest DFT free energy,
the phase field crystal (PFC) class \cite{pfc04,*pfcdft07}.

The dimensionless Helmholtz potential of a 
two component PFC system can be written \cite{chanphd,*vpfc09}
\begin{equation}
F=\int d\vec{r} \left[ f_{A}+f_{B}+f_{AB} \right]
\label{pfcfreetot}
\end{equation}
where
\[
f_{i}=\frac{n_i}{2} \left[ B_i^{\ell}+(q_i^2+\nabla^2)^2 \right] n_{i} - 
\frac{w_i}{3}n_{i}^3 + \frac{u_i}{4}n_{i}^4 + H_{i}(|n_{i}|^3-n_{i}^3)
\]
and
\[
f_{AB}=\frac{n_A}{2}(q_{AB}^2 + \nabla^2)^2 n_B + 
\frac{\lambda_1}{2} n_A^2 n_B^2 + \frac{\lambda_2}{2}(\nabla \delta c)^2.
\]
In this notation $i=A$ or $B$, $n_i=n_i(\vec{r},t)+\bar{n}_i$ is the scaled
time averaged number density of $i$ particles, $\bar{n}_i$ is the 
species average
number density, $B_i^{\ell}$ is related to the liquid bulk modulus,
$q_i$ sets the equilibrium distance between particles of the same species,
$q_{AB}$ sets that between $A$ and $B$ particles,
and $w_i$, $u_i$, $H_i$, $\lambda_1$, and $\lambda_2$ are constants 
(see Refs.~\cite{pfc04,*pfcdft07} for further discussion of how these 
parameters relate to material properties).
The terms multiplied by $H_i$ discourage $n_i < 0$
and are the distinguishing feature of the Vacancy or VPFC model
\cite{chanphd,*vpfc09}.
A hard $n_i \ge 0$ cutoff enforces the physical interpretation of $n_i$
as a number density and in doing so
produces a range of nonlinear responses.
The resulting solutions take the form of interacting time-averaged
density peaks, with local regions of $n_i \simeq 0$ representing unoccupied, 
or vacancy, sites. In terms of glass formation,
one can obtain qualitatively similar results whether $H_i=0$ or
$H_i \ne 0$, but only the $H_i \ne 0$ case is presented in the following.

    The simplest dynamics conserving $n_i$ may be written
\begin{equation}
\frac{\partial n_i}{\partial t}= 
\nabla^2 \frac{\delta F}{\delta n_i} + \sqrt{D_i}\eta_i
\label{pfcdyn1}
\end{equation}
where $t$ is dimensionless time,
$D_i \sim T$, and
$\eta_i$ is a gaussian stochastic noise variable with 
$\langle\eta_i(\vec{r}_1,t_1)\eta_i(\vec{r}_2,t_2)\rangle
=\nabla\cdot\nabla\delta(\vec{r}_1-\vec{r}_2)\delta(t_1-t_2)$.
A second option is the overdamped equation of DDFT,
\begin{equation}
\frac{\partial n_i}{\partial t}= 
\nabla \cdot \left( n(\vec{r},t) \nabla \frac{\delta F}{\delta n_i} \right)
+\sqrt{D_i}\nu_i
\label{pfcdyn3}
\end{equation}
where 
$n(\vec{r},t)$ is generally set to $n_i(\vec{r},t)$ and
$\langle\nu_i(\vec{r}_1,t_1)\nu_i(\vec{r}_2,t_2)\rangle
=\nabla \cdot \nabla \left[n(\vec{r},t)\delta(\vec{r}_1-\vec{r}_2)
\delta(t_1-t_2)\right]$.
A third equation reintroduces some of the faster
dynamics by also including an inertial or wave-like term,
\begin{equation}
\frac{\partial^2 n_i}{\partial t^2}+
\beta_i\frac{\partial n_i}{\partial t}= 
\alpha_i^2 \nabla \cdot \left( n(\vec{r},t) \nabla \frac{\delta F}{\delta n_i} 
\right) + \sqrt{D_i}\nu_i
\label{pfcdyn2}
\end{equation}
where $\alpha_i$ and $\beta_i$ are constants \cite{mpfc}.

Previous PFC simulations indicate that
Eq.~(\ref{pfcdyn1}) supports metastable glassy states but in general 
produces a discontinuous, nucleation driven 
liquid to glass transition \cite{pfcglass08}.
Recent analyses of Eqs.~(\ref{pfcdyn3}) and (\ref{pfcdyn2})
suggest that both may recover the class of MCT equations for the liquid
dynamic correlators that
successfully describe a wide range of glass forming behaviors
\cite{ABL06,*kimkawasaki07,*archer09}.
Here we numerically investigate, without approximation,
Eq.~(\ref{pfcdyn2}) with $n(\vec{r},t)=1$, where
the inclusion of stochastic noise implies a time-averaged rather than
ensemble-averaged interpretation of DDFT 
\cite{archer04,*kawasaki06}.
Equilibrium liquid states at high $D_i$ were quenched by lowering
the stochastic noise amplitude $T=T_0 D_i$ at a rate $\dot{T}$, 
and the freezing transition was
analyzed for onset of vitrification or crystallization.

\begin{figure}[btp]
 \raggedright
 \subfigure{\includegraphics*[width=0.235\textwidth,height=0.175\textwidth,trim=0 0 0 0]{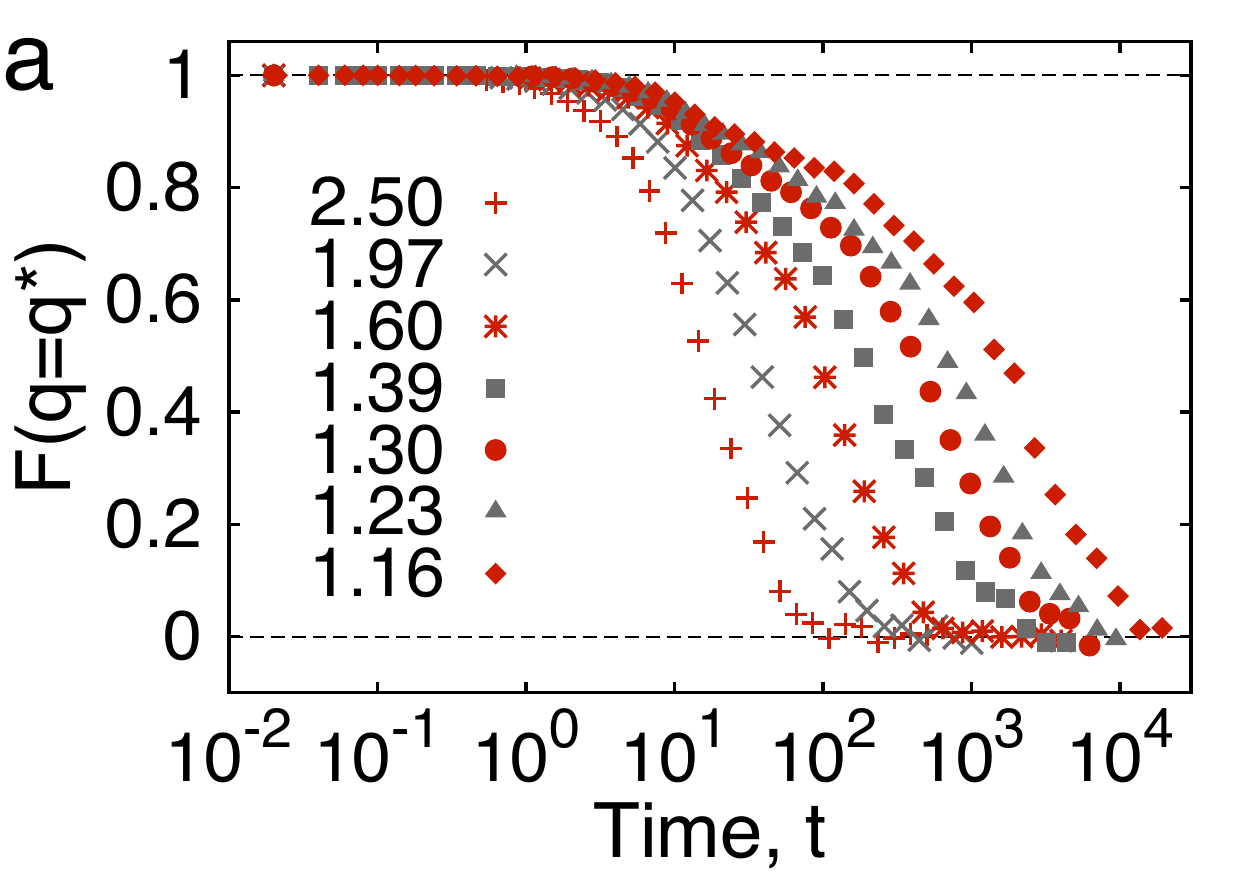}}%
 \hspace{.1 cm}
 \subfigure{\includegraphics*[width=0.235\textwidth,height=0.175\textwidth]{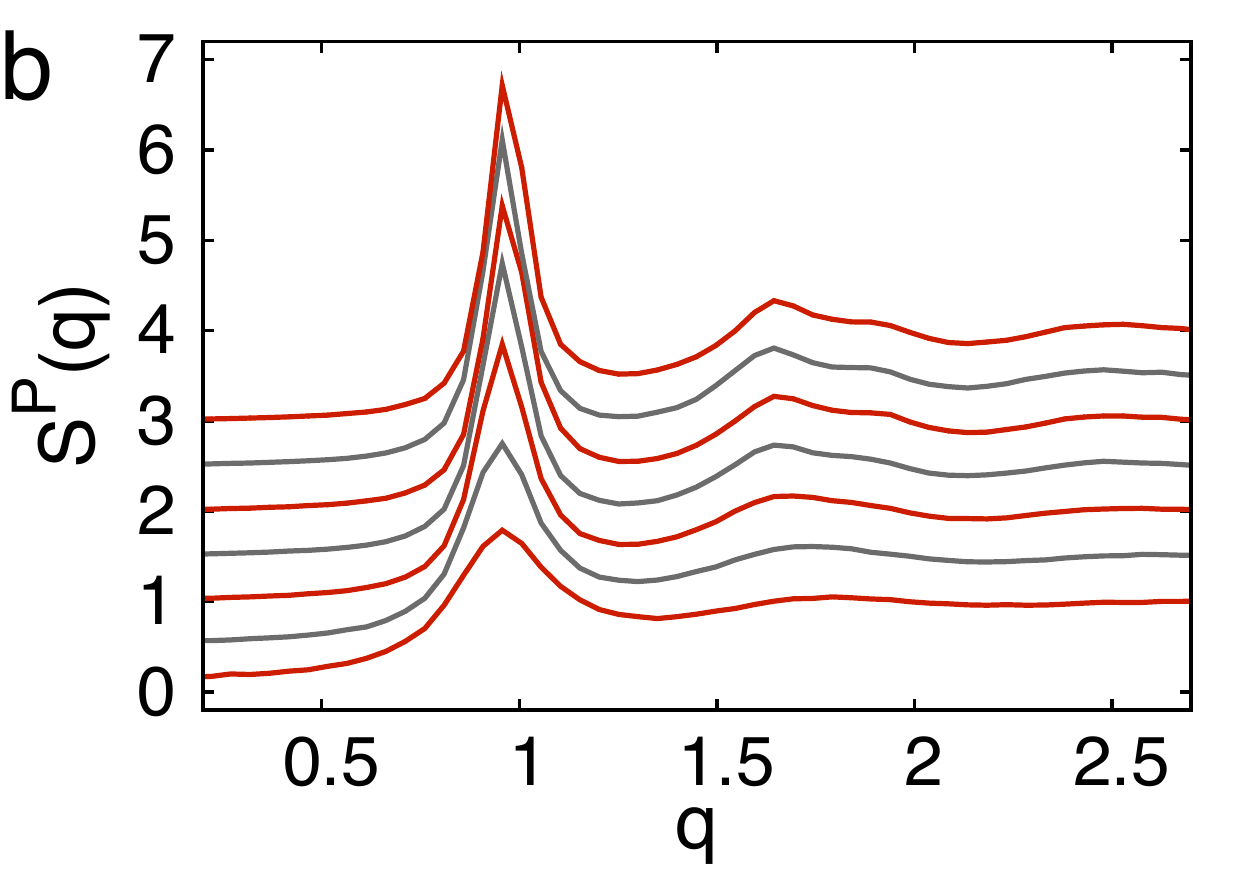}}\\
 \vspace{-.1 cm}
 \hspace{-.15 cm}
 \subfigure{\includegraphics*[width=0.235\textwidth,height=0.175\textwidth,trim=0 0 0 0,clip]{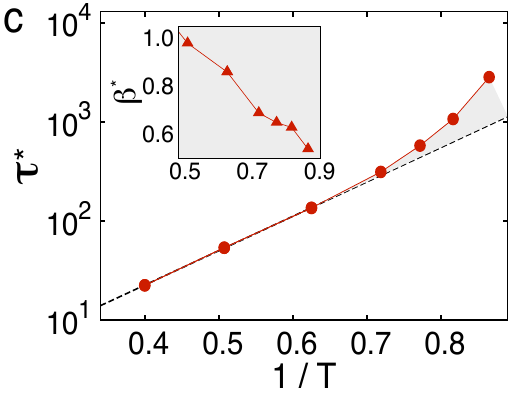}}%
 \hspace{.1 cm}
 \subfigure{\includegraphics*[width=0.235\textwidth,height=0.175\textwidth]{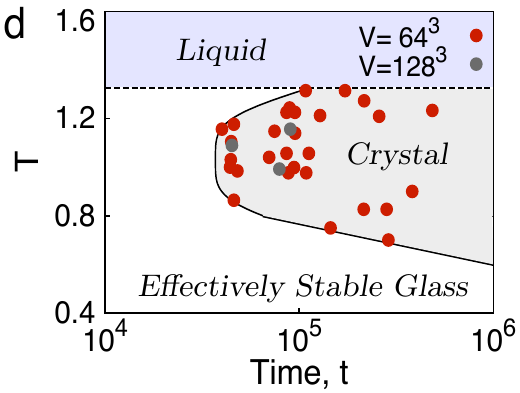}}
\vspace{-.6 cm}
\caption{\label{puredat} (color online).
Glass formation and crystallization in the monatomic VPFC model.
(a) $F(q^*,t)$ at
various $T$ where $q^*$ corresponds to the first peak maximum in $S^P(q)$,
(b) $S^P(q)$ at same $T$ as in (a), 
offset vertically by $0.5n$ with $n=0,1,...$,
(c) Arrhenius plot of $\tau^*$ from $F(q^*,t)$,
Inset: stretching exponent $\beta^*$ from fit to
$F(q^*,t)=\exp{[-(t/\tau^*)^{\beta^*}]}$,
(d) $TTT$ diagram: quenches from $T=1.6$ at various $\dot{T}$, 
points denote where crystallization occurred.
$\bar{n}_A=0.15$, $B_A^{\ell}=-0.9$, $q_A=1$, $w_A=0$, $u_A=1$, $H_A=1500$,
$T_0=1000$, $\alpha_A=1$, $\beta_A=0.01$, $f_B=f_{AB}=0$,
$\Delta x=1.0$, $\Delta t=0.02$, $V=128^3$.
}
\end{figure}

We begin with results for monatomic systems, outlined in Fig.\ \ref{puredat}.
For $T \gtrsim 1.6$
the structure and dynamics are those of a normal liquid. 
The measured intermediate scattering functions
($F_{ij}(q,t)=\langle \delta n_i(q,0) \delta n^*_j(q,t) \rangle / F_{ij}(q,0)$)
decay exponentially,
the corresponding average relaxation times
show an Arrhenius $T$ dependence, 
and the structure factors
are characteristic of an equilibrium liquid state.
The function $S^P(q)$ quantifies the structural correlations of the localized
peaks in the density field.
We define $S^P_{ij}(q)=\langle \delta n^P_i(q) \delta n^{P*}_j(q') \rangle$, 
where $\delta n^P_i(r)$ is a binary map of
the positions of the local number density peaks.

Below $T \simeq 1.6$, the liquid begins to show signs of nonequilibrium
behavior and the onset of glass formation. 
$F(q,t)$ becomes increasingly stretched and begins to exhibit a shoulder,
the average relaxation
time briefly begins to grow with a super-Arrhenius $T$ dependence, and
a split second peak emerges in $S^P(q)$.
But signs of glass formation persist only to 
the freezing temperature of the crystal, $T_f$.
Below this point crystallization 
interrupts the apparent glass transition 
unless the liquid is rapidly quenched well below $T_f$. 
The time-temperature-transformation ($TTT$) diagram
shown in Fig.~\ref{puredat}d demonstrates this behavior.
The profile of the nose feature 
is typical for
materials with relatively marginal
glass forming ability, such as metallic glasses.
Since long-lived glassy states are not supported in the region 
$0.6 \lesssim T \lesssim T_f$, one cannot study a gradual dynamic transition
from liquid to glass. This behavior is expected for simple monatomic 
systems.

\begin{figure}[btp]
 \centering
 \subfigure{\includegraphics*[width=0.235\textwidth,height=0.175\textwidth]
  {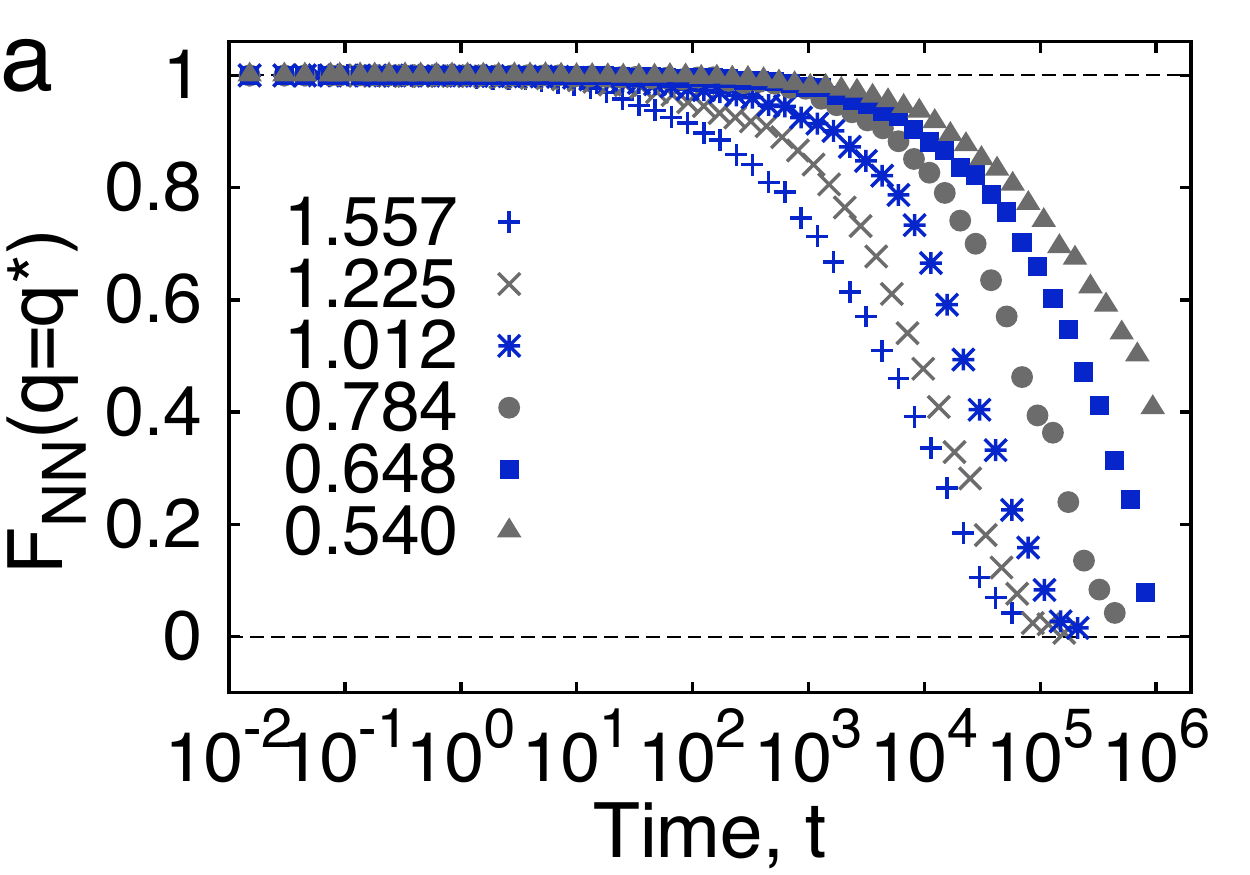}}%
\hspace{.1 cm}
 \subfigure{\includegraphics*[width=0.235\textwidth,height=0.175\textwidth]
  {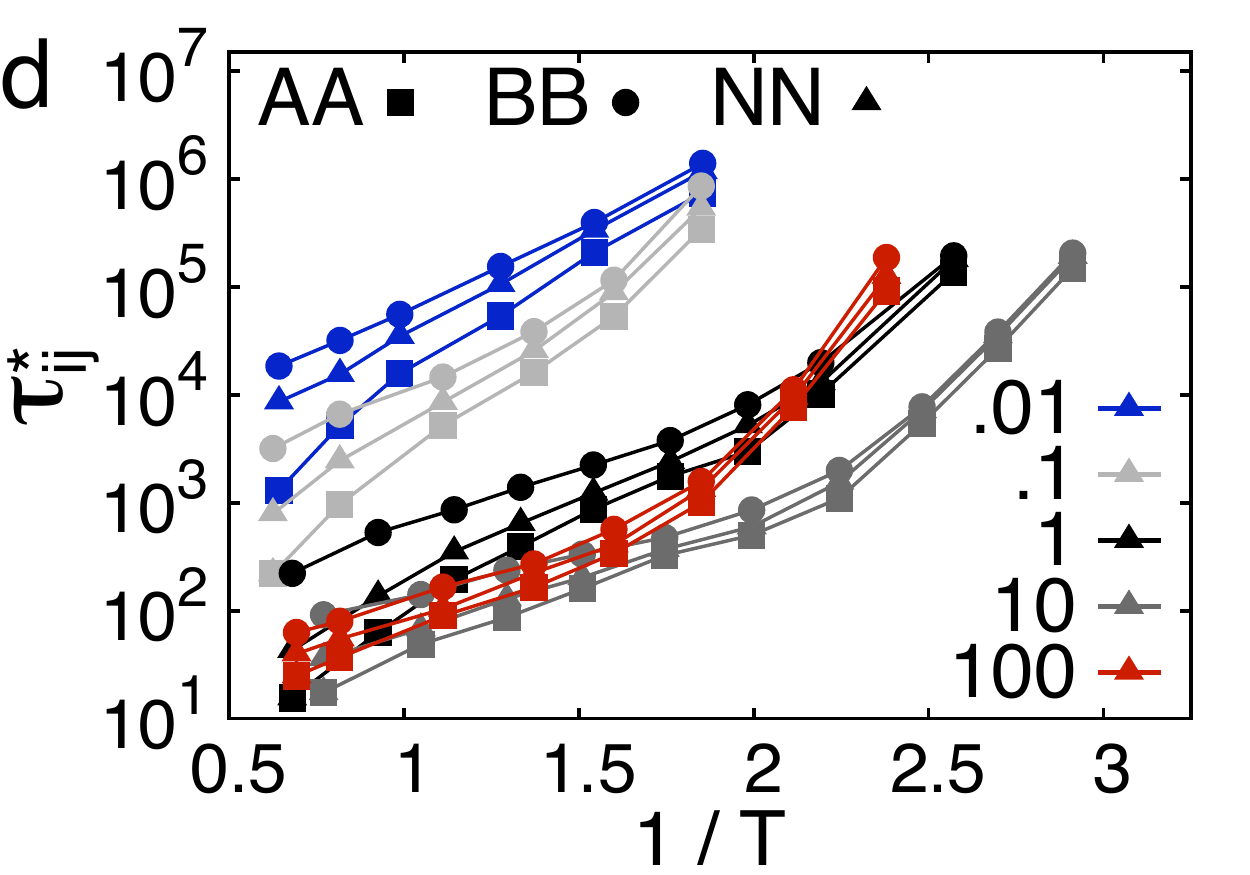}}\\
\vspace{-.3 cm}
 \subfigure{\includegraphics*[width=0.235\textwidth,height=0.175\textwidth]
  {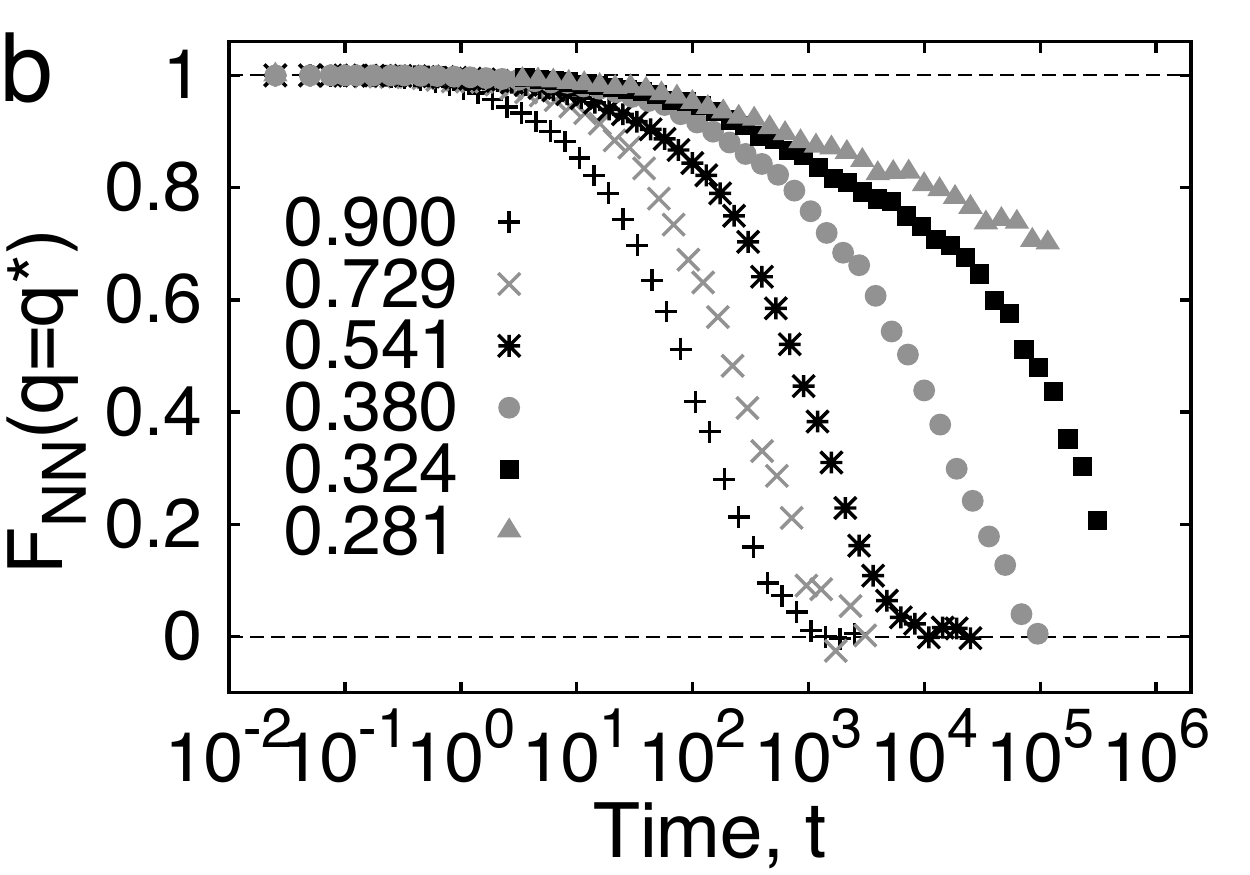}}%
\hspace{.1 cm}
 \subfigure{\includegraphics*[width=0.235\textwidth,height=0.175\textwidth]
  {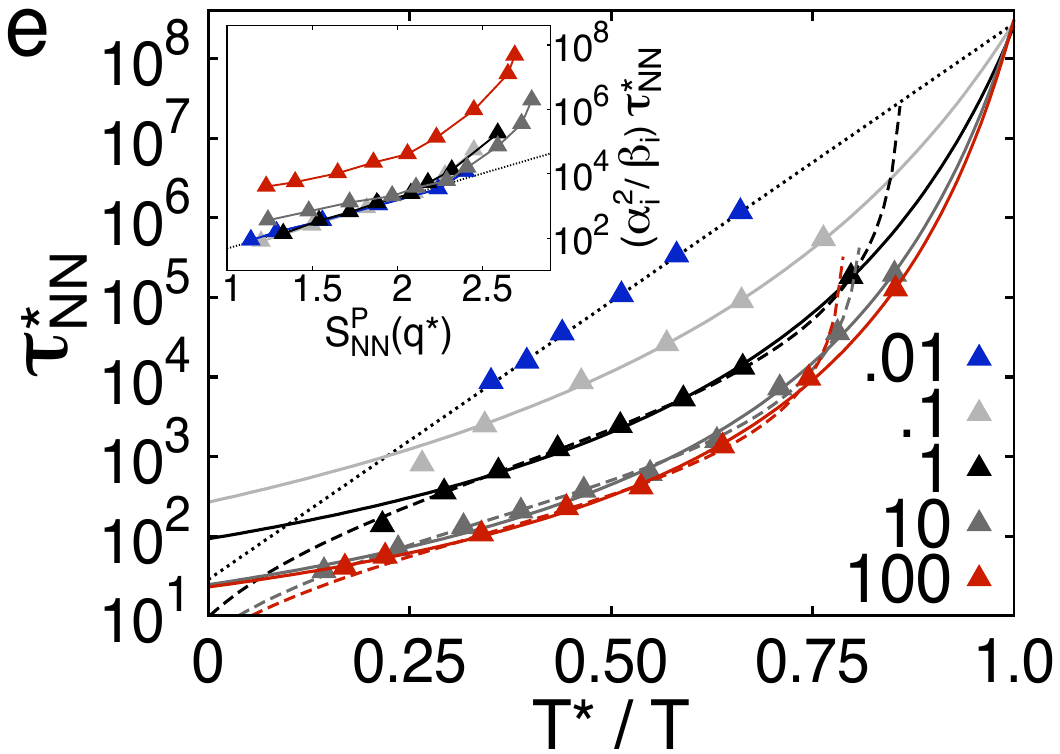}}\\
\vspace{-.3 cm}
 \subfigure{\includegraphics*[width=0.235\textwidth,height=0.175\textwidth]
  {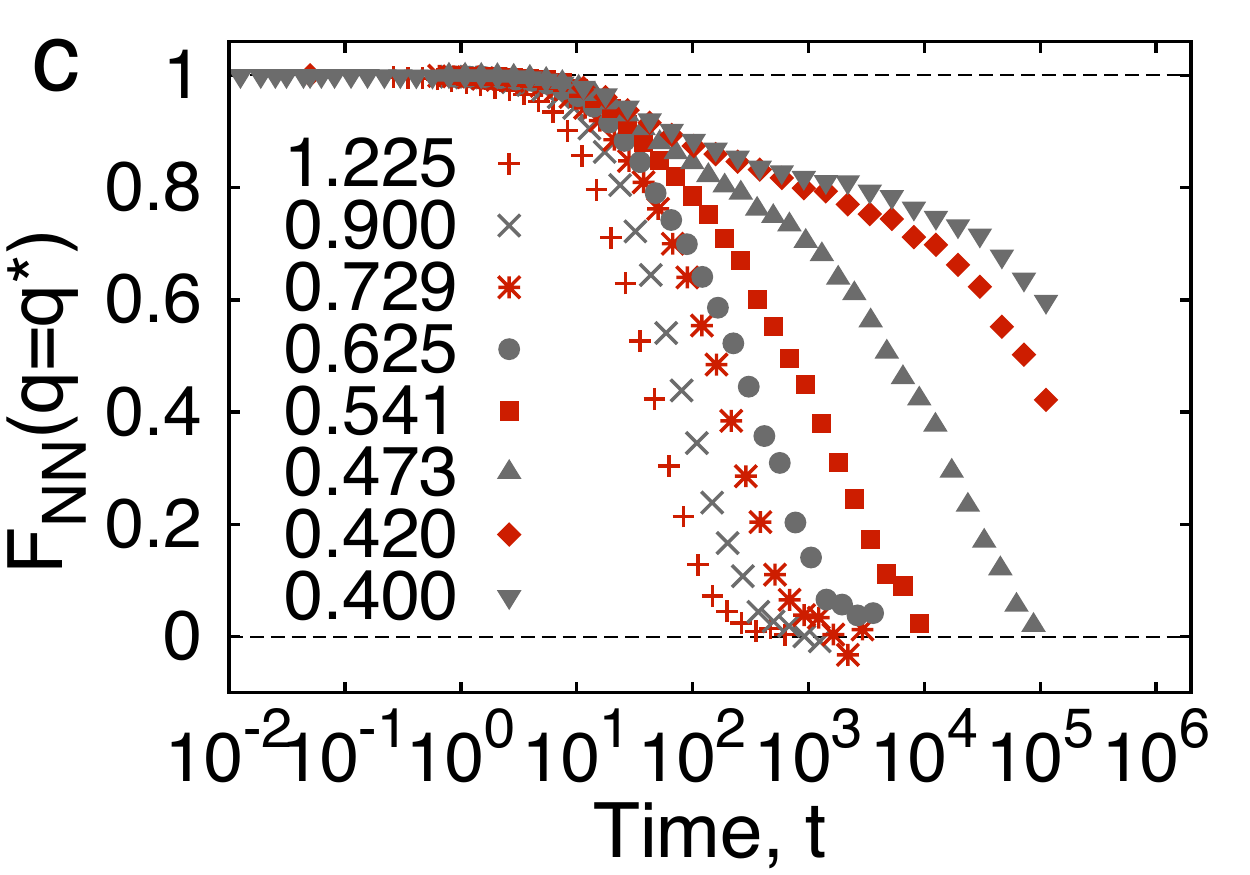}}%
\hspace{.1 cm}
 \subfigure{\includegraphics*[width=0.235\textwidth,height=0.175\textwidth]
  {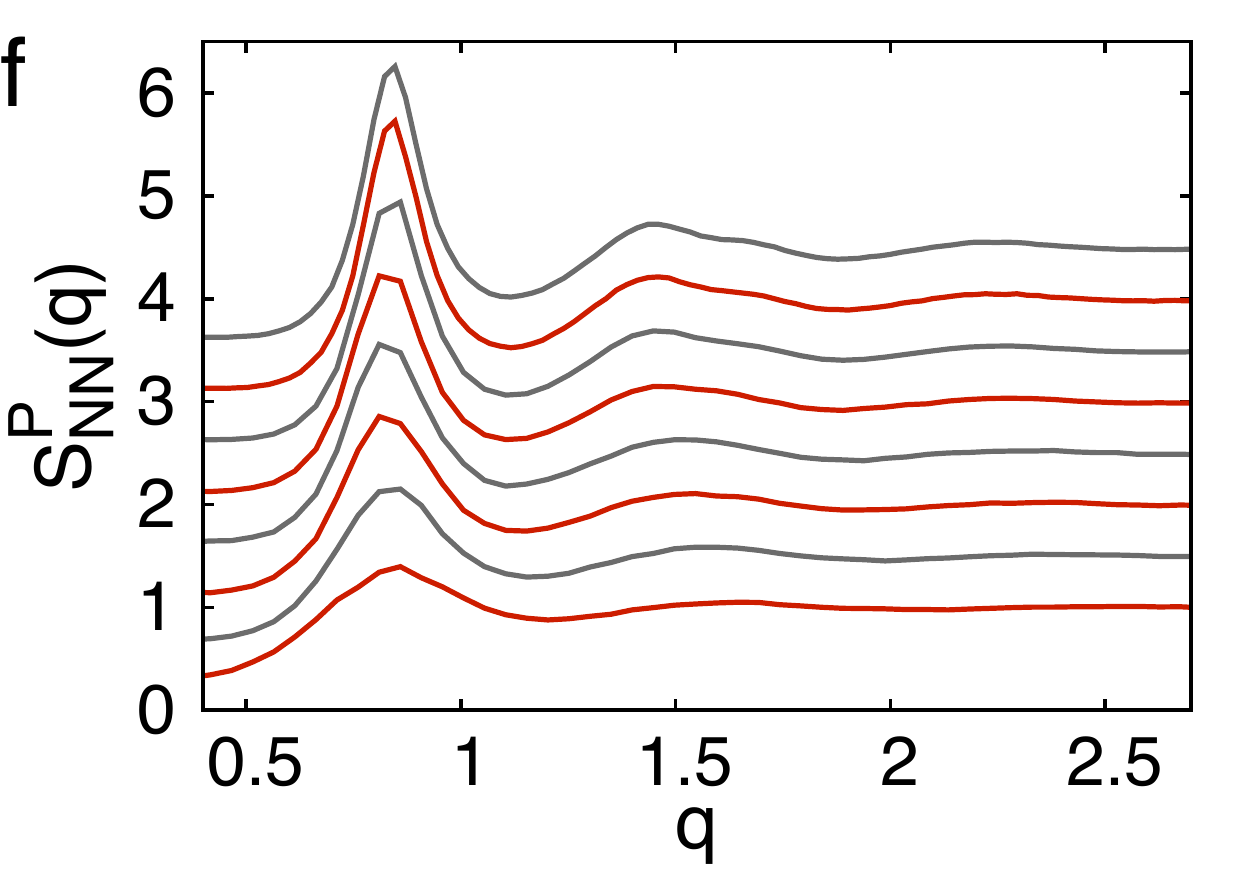}}%
\vspace{-.35 cm}
\caption{\label{dynrange} (color online).
Binary VPFC results for various damping conditions.
$F_{NN}(q^*,t)$ at various $T$
shown for $\alpha_i/\beta_i$ of (a) $0.01$, (b) $1$, and (c) $100$.
(d) Arrhenius plot of the structural relaxation times $\tau^*_{ij}$,
(e) the same data ($NN$ only) 
scaled to clarify strong vs.~fragile behaviors,
with Vogel-Fulcher fits shown as solid lines and power law fits
as dashed lines. 
Inset: $(\alpha^2_i/\beta_i)\tau^*_{NN}$ vs.~$S^P_{NN}(q^*)$, 
demonstrating deviation from scaling
when inertial effects become large.
(f) $S^P_{NN}(q)$ for $\alpha_i/\beta_i=100$.
$\bar{n}_A=\bar{n}_B=0.075$, $B_i^{\ell}=-0.9$, $q_A=1$, $q_B=0.8$,
$w_i=0$, $u_i=1$, $H_i=1500$, $T_0=1000$, 
$q_{AB}=8/9$, $\lambda_1=100$, $\lambda_2=0$, 
$\Delta x=1.0$, $\Delta t=0.025$, $V=64^3$, $128^3$, or $256^3$.
}
\end{figure}

Thus we proceed to binary liquids and outline in Fig.~\ref{dynrange}
the qualitative behavior of one such model system for a range of dynamic
conditions, from highly underdamped ($\alpha_i/\beta_i=100$) to 
highly overdamped ($\alpha_i/\beta_i=0.01$).
The chosen model contains equal number densities
of $A$ and $B$ atoms ($\bar{n}_A=\bar{n}_B=0.075$), 
and the equilibrium spacing of $A$ atoms is $20\%$ 
smaller than that of $B$ atoms ($R_A/R_B=q_B/q_A=0.8$).
Only the $NN$ correlations are plotted in Figs.~\ref{dynrange}a-c,
where $N$ denotes the full density field $n_A+n_B$.
When damping dominates,
an effectively stable glass
with dynamics resembling those of a strong glass former is generated.
The dynamic correlators are
generally best fit as a single exponential decay for all accessible $T$,
with increased stretching as $T$ is lowered, but any plateaus
are absent or ill-defined in the $\alpha_i/\beta_i \lesssim 1$ data.
The relaxation times exhibit a nearly Arrhenius $T$-dependence
over the entire accessible $T$-range.
 
At the opposite extreme, when inertia dominates,
a transition with dynamics characteristic of fragile liquids is generated.
The dynamic correlators show both stretching and clear
plateauing as $T$ is lowered, and
the divergence of the relaxation time is well fit by the Vogel-Fulcher
form ($\tau=A\exp{[B/(T-T_0)]}$).
This divergence becomes increasingly super-Arrhenius at higher $T$ as
$\alpha_i/\beta_i$ grows.
The underdamped transition at this level of detail
qualitatively resembles that described by MCT.

The fragility of the PFC liquid therefore appears to be
strongly linked with the balance of inertial and damping terms 
in Eq.~(\ref{pfcdyn2}), $\alpha_i/\beta_i$.
The degree of fragility is in turn correlated with
the nominal spatial extent of cooperative dynamic behavior, 
which is set by an inherent length scale
associated with the inertial term. 
This term generates wave modes which propagate
over a fixed length scale in a crystal before being damped, and
the resulting dynamic correlation length 
follows $\xi^{\rm crystal}_D \sim \alpha_i/\beta_i$ \cite{mpfc}.
In a normal liquid these correlations are largely suppressed by 
the low density and weak structural correlations, so that 
$\xi^{\rm liquid}_D \ll \xi^{\rm crystal}_D$. 
But with greater supercooling, as the
system becomes increasingly dense and solid-like,
the inertial correlations survive over length scales which likely approach
$\xi^{\rm crystal}_D$. 
Roughly, 
$\xi^{\rm liquid}_D \rightarrow \xi^{\rm crystal}_D \sim \alpha_i/\beta_i$
as $T \rightarrow T_g$ 
\cite{percnote}.

\begin{figure}[btp]
 \centering
\vspace{0 cm}
\hspace{-.1 cm}
 \subfigure{\includegraphics*[width=0.15\textwidth,trim=60 0 60 50]{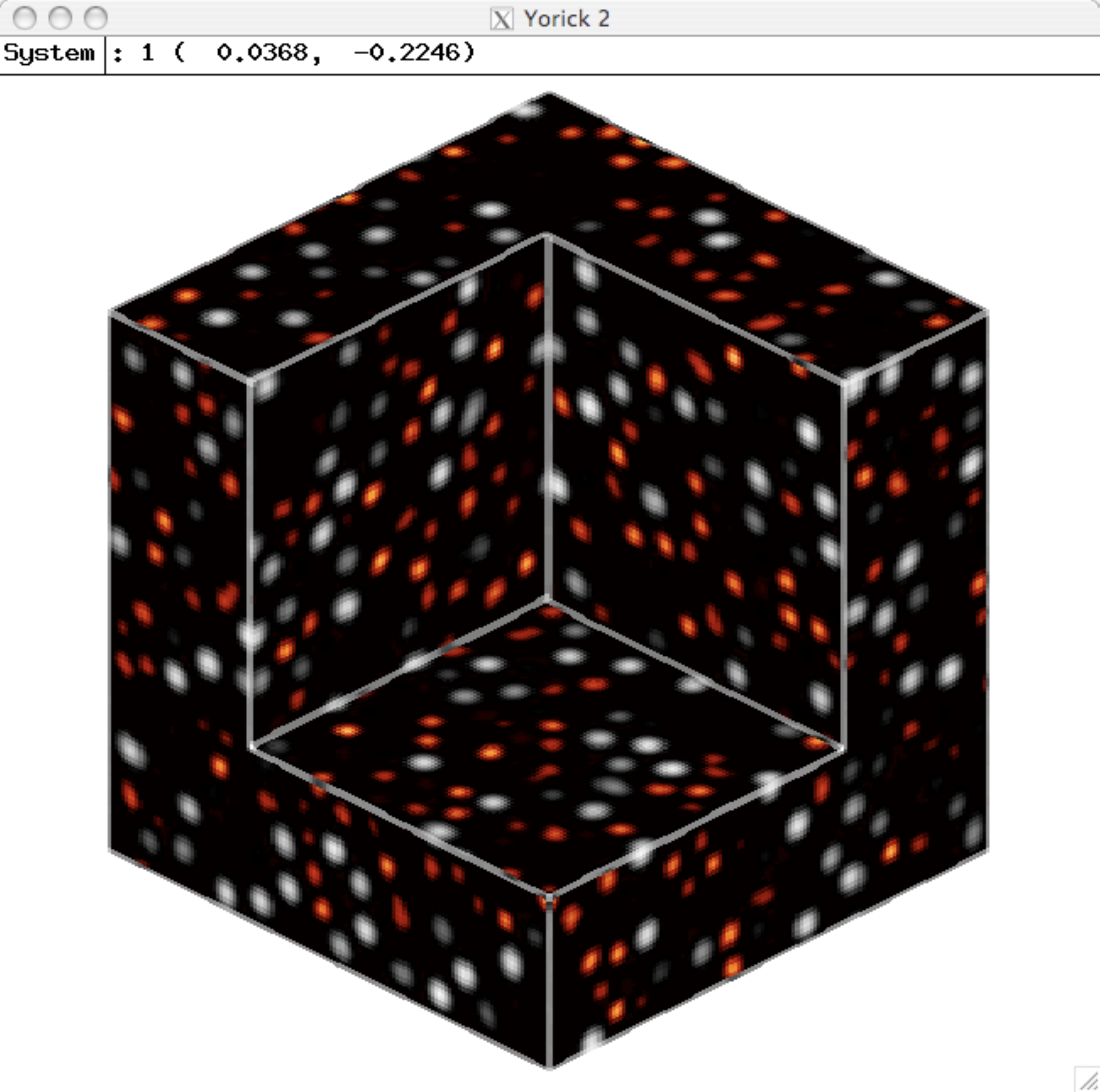}\put(-54,94){\underline{$t=\tau^*_{NN}$}}}
\hspace{.0 cm}
 \subfigure{\includegraphics*[width=0.15\textwidth,trim=60 0 60 50]{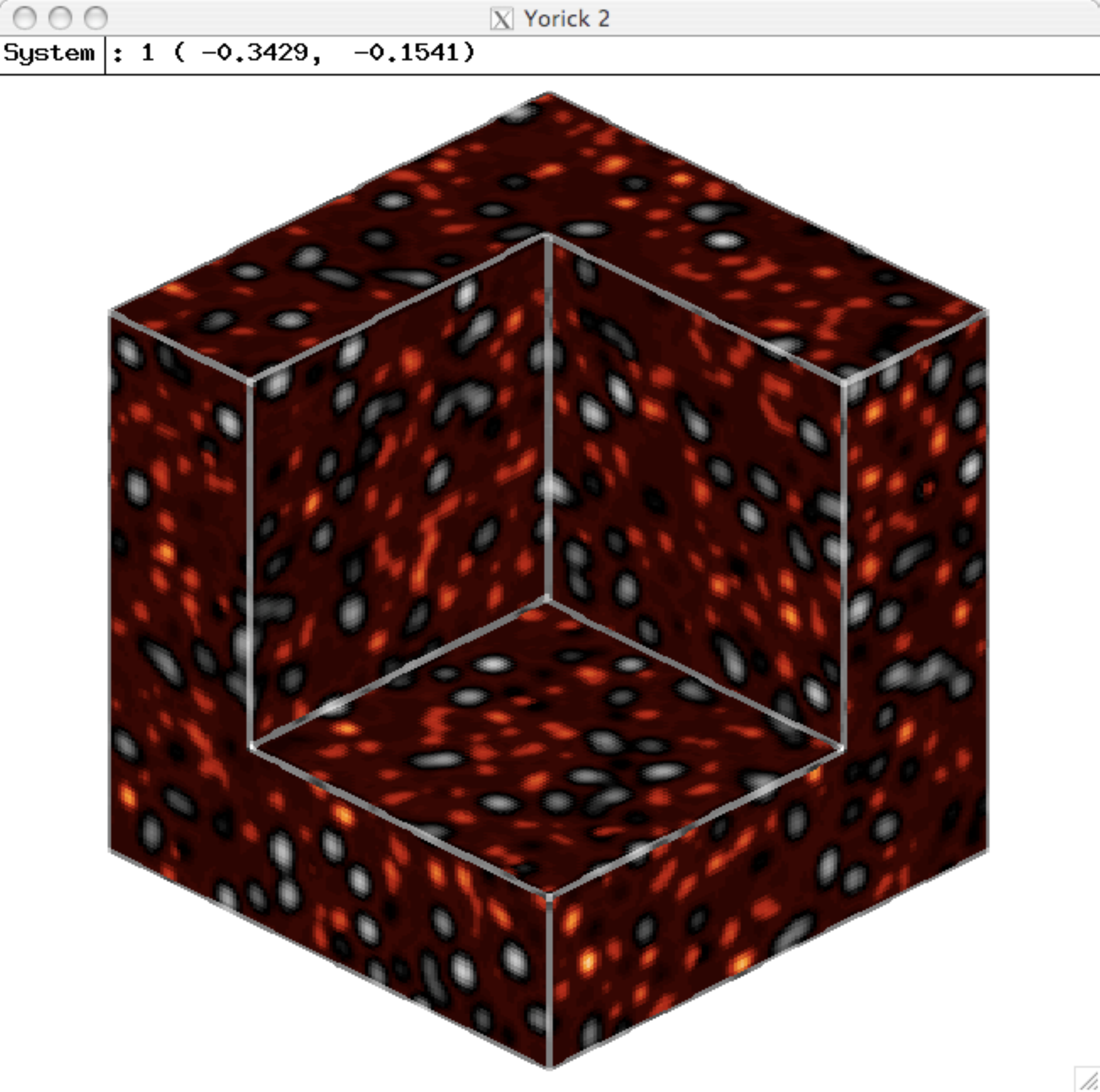}\put(-61,94){\underline{$t=5.7\tau^*_{NN}$}}}
\hspace{.0 cm}
 \subfigure{\includegraphics*[width=0.15\textwidth,trim=60 0 60 50]{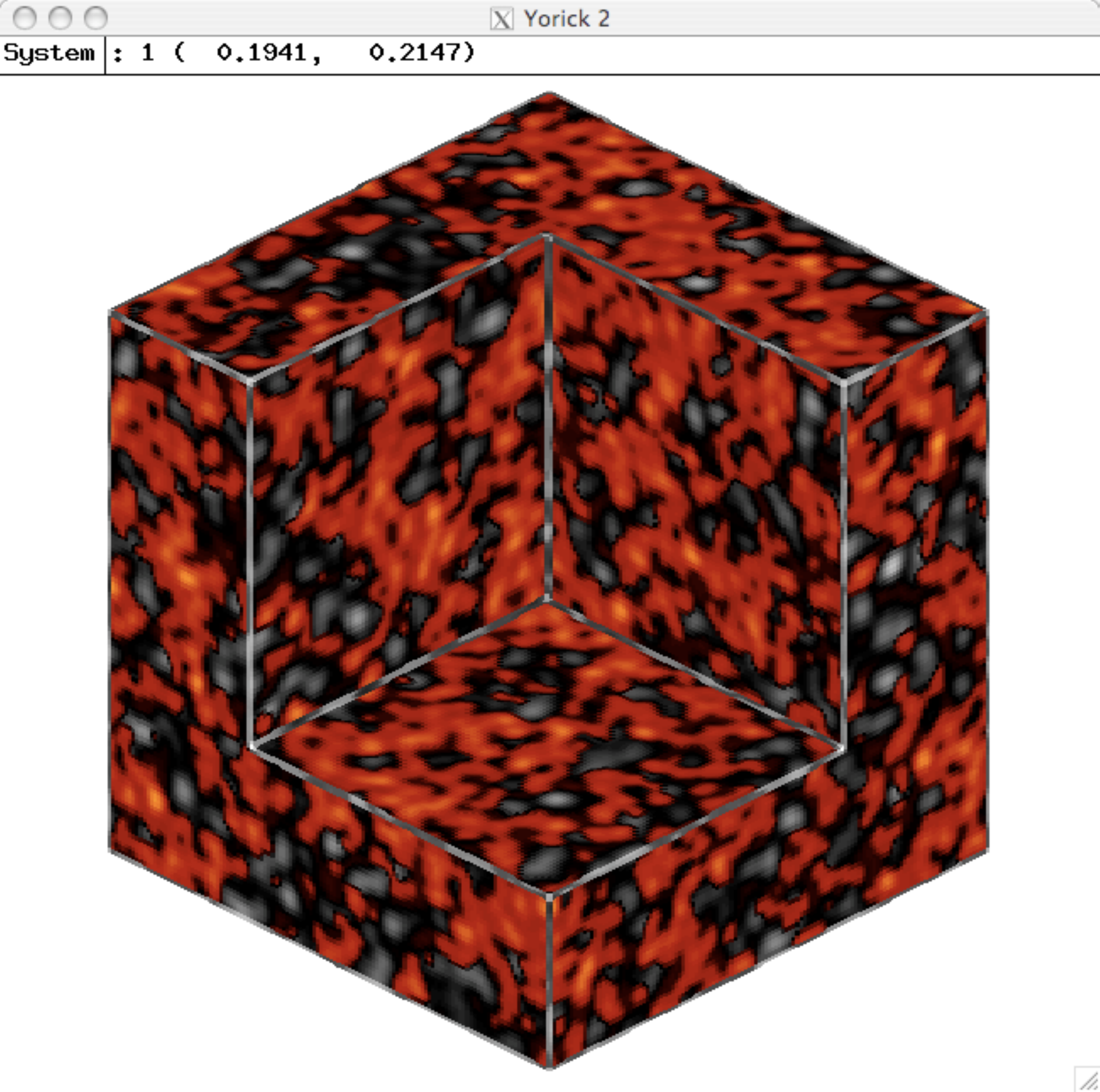}\put(-64,94){\underline{$t=38.5\tau^*_{NN}$}}}\\
\vspace{-.1 cm}
\hspace{-.1 cm}
 \subfigure{\includegraphics*[width=0.15\textwidth,trim=60 0 60 50]{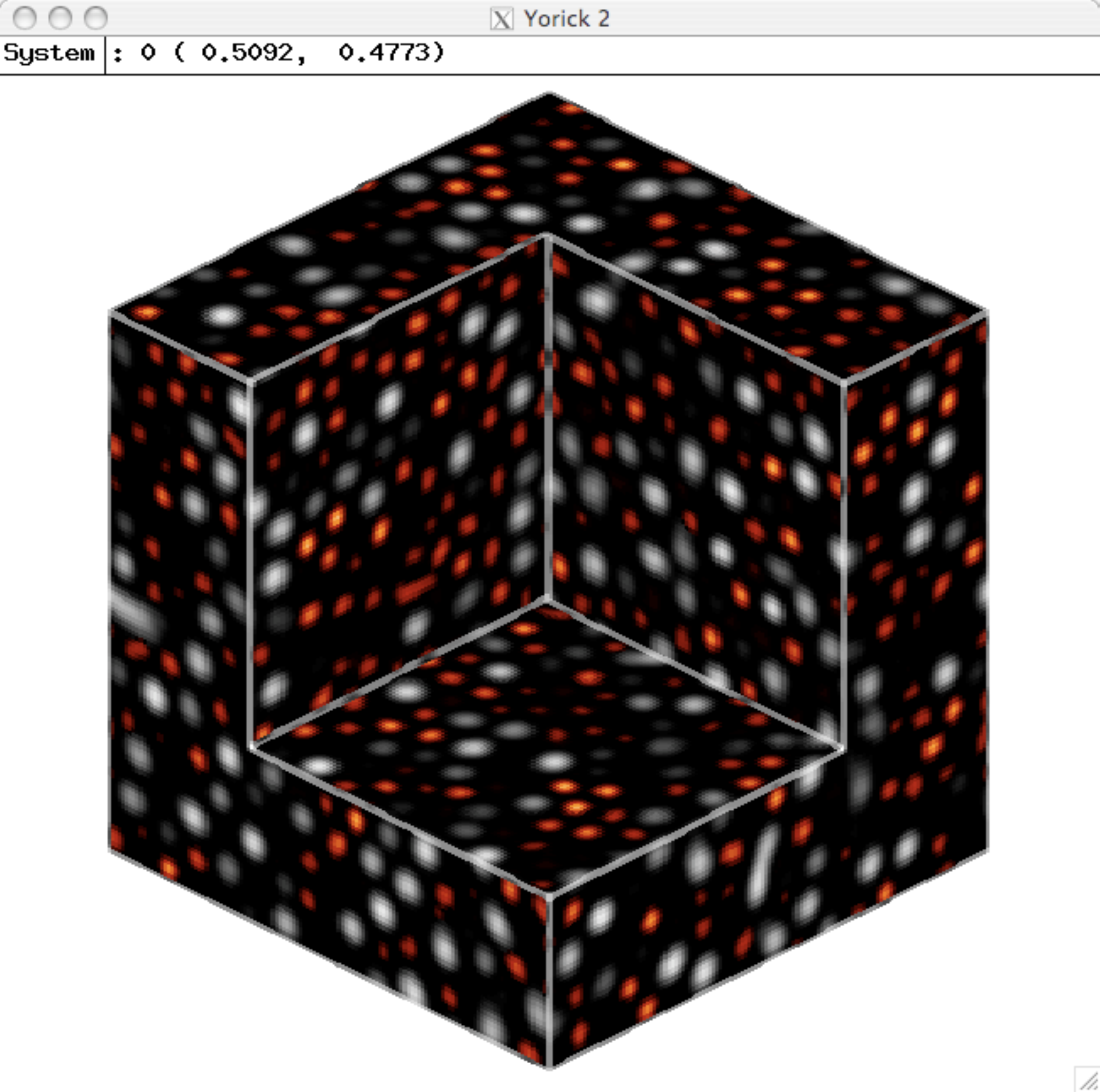}}
\hspace{.0 cm}
 \subfigure{\includegraphics*[width=0.15\textwidth,trim=60 0 60 50]{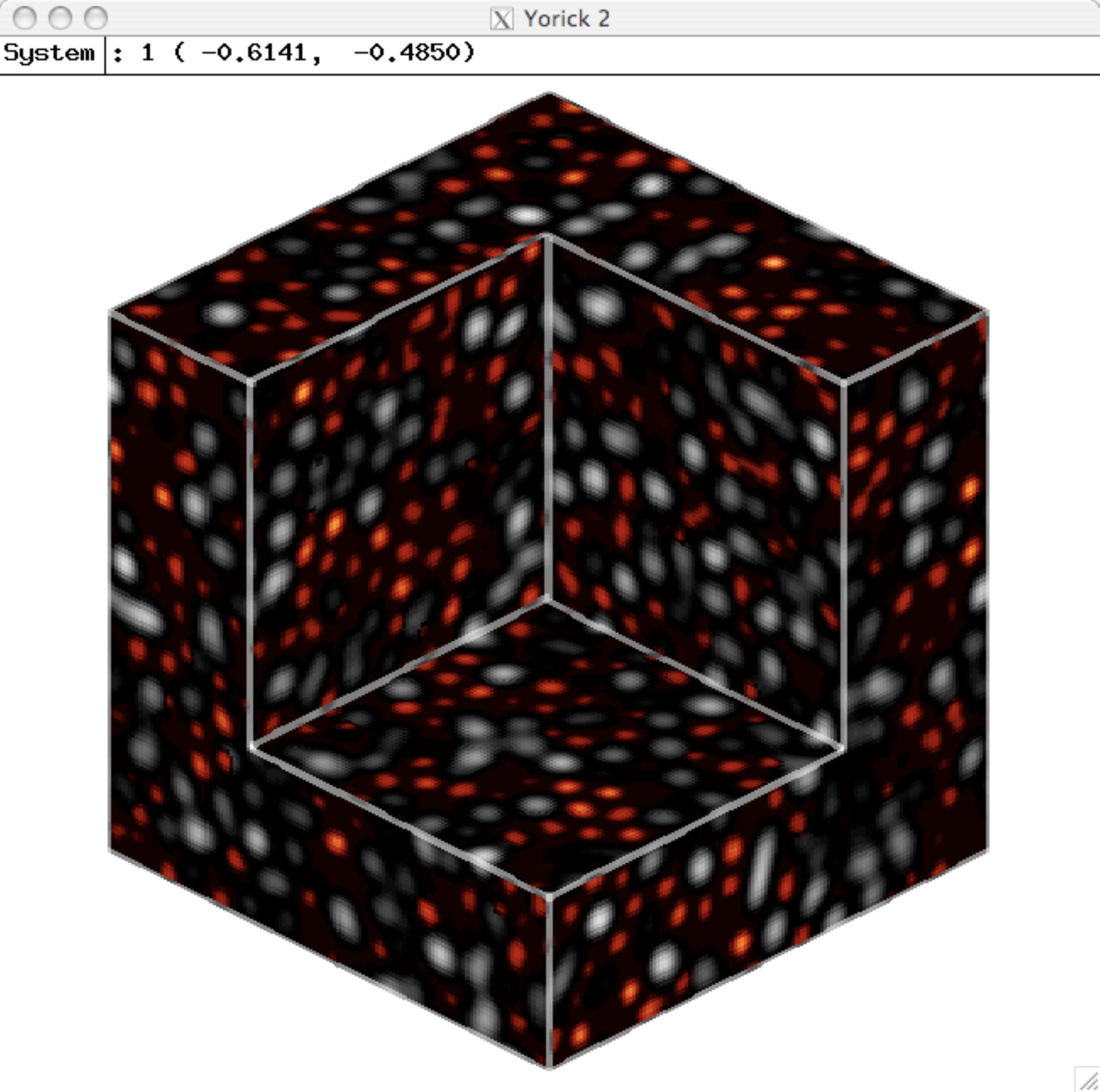}}
\hspace{.0 cm}
 \subfigure{\includegraphics*[width=0.15\textwidth,trim=60 0 60 50]{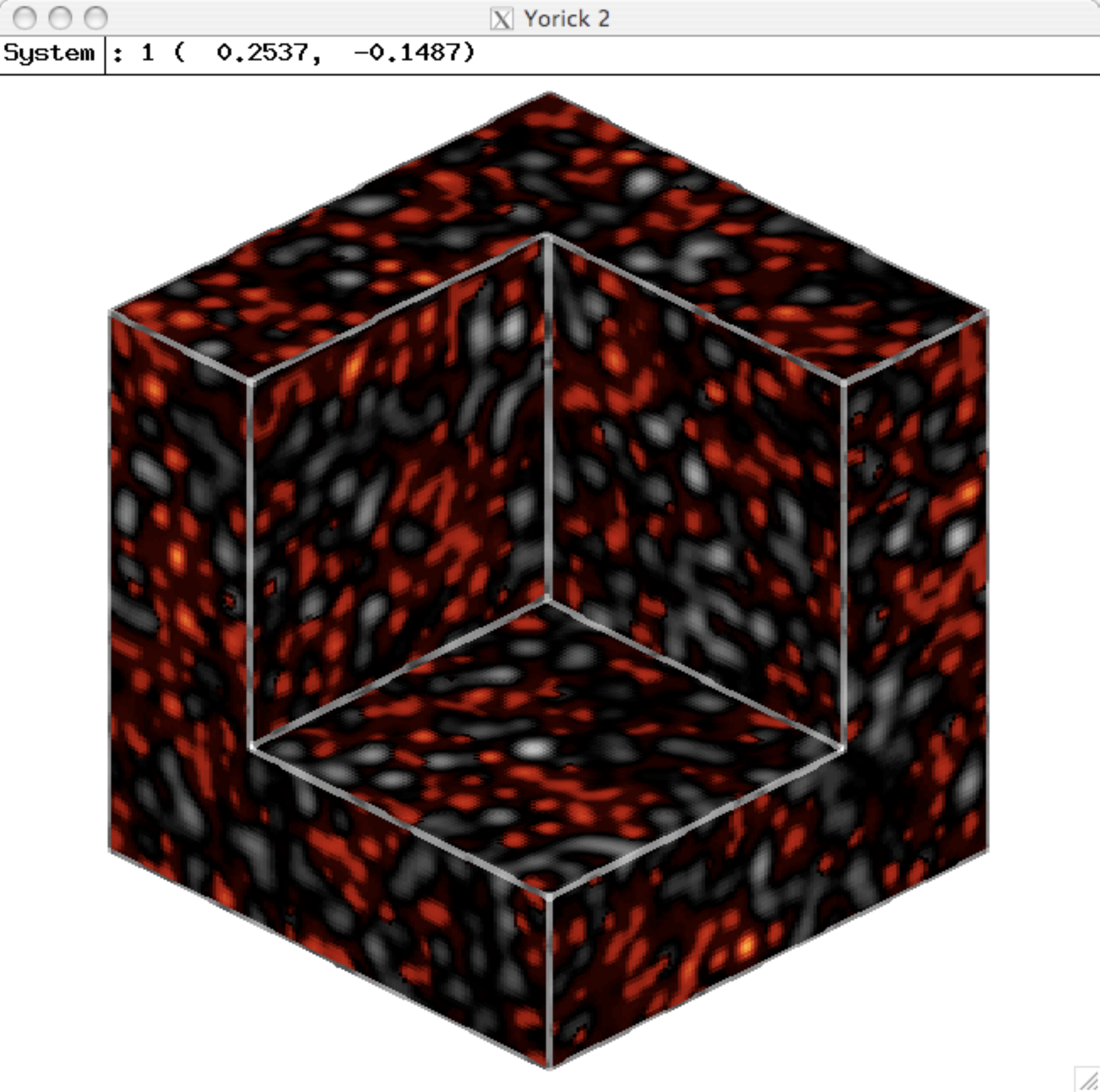}}\\
\vspace{-.35 cm}
\hspace{-.031 cm}
\vspace{0. cm}
 \subfigure{\includegraphics*[width=0.15\textwidth,trim=60 0 60 50]{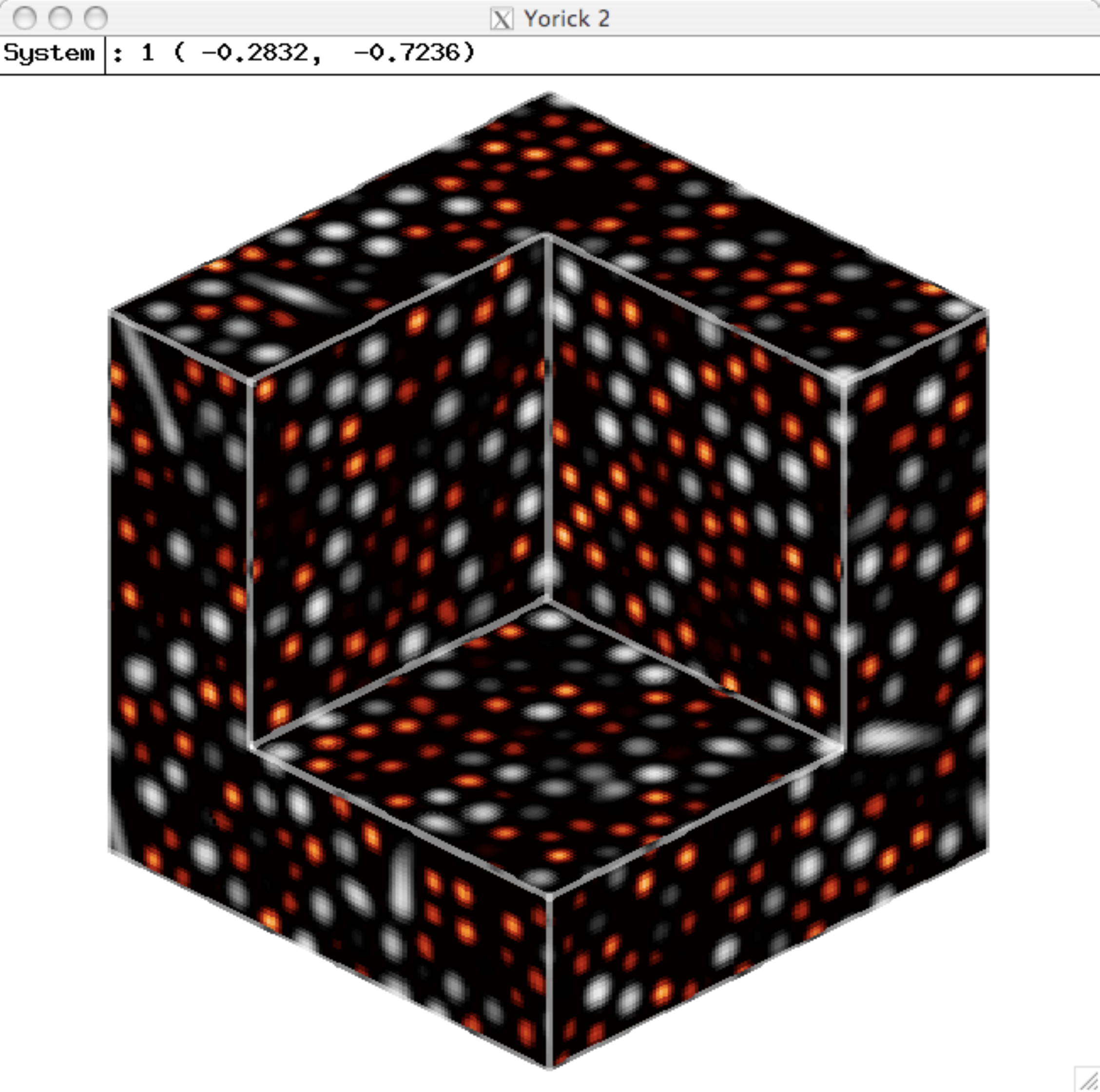}}
\hspace{.0 cm}
 \subfigure{\includegraphics*[width=0.15\textwidth,trim=60 0 60 50]{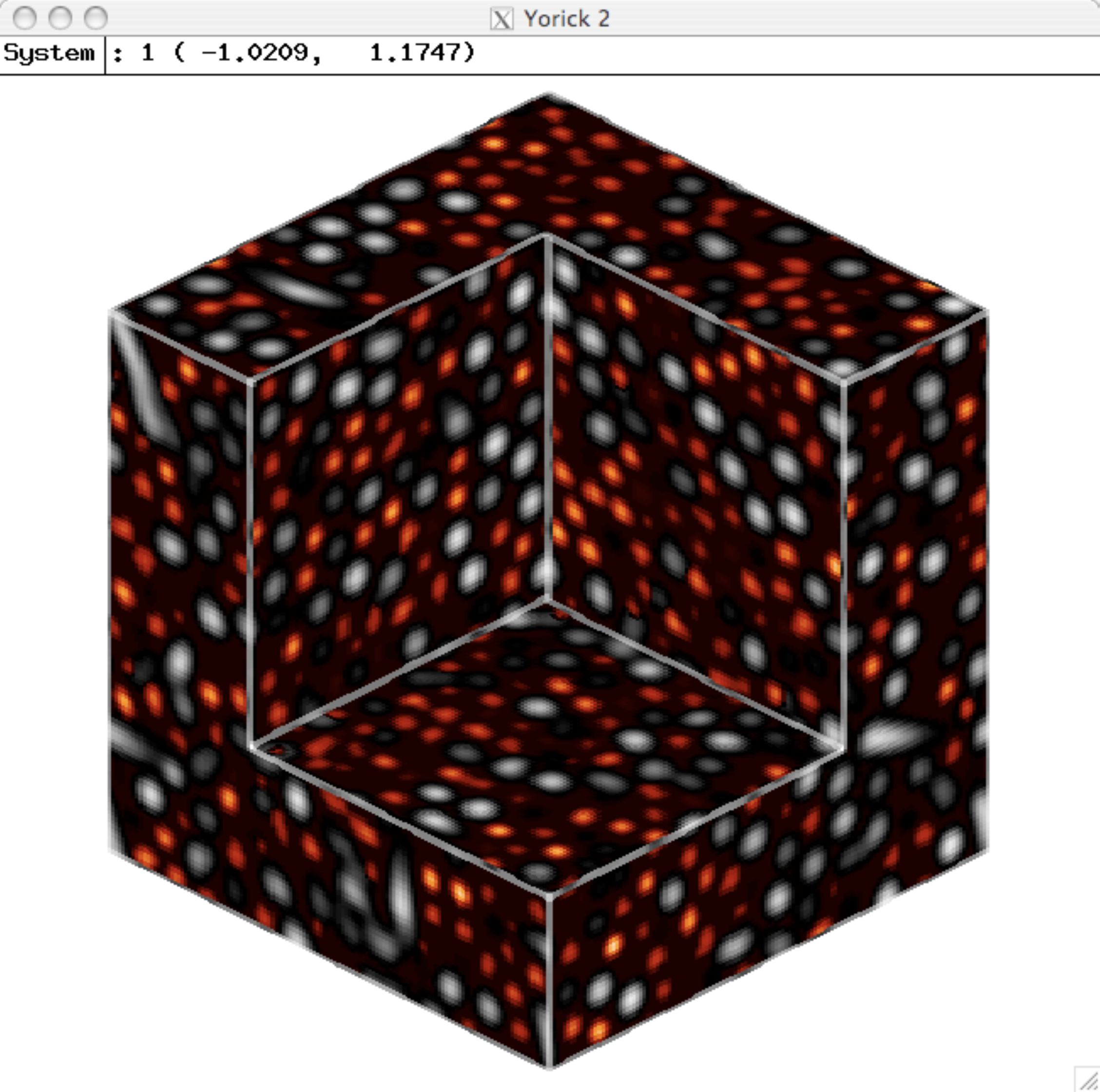}}
\hspace{-.05 cm}
 \subfigure{\includegraphics*[width=0.155\textwidth,height=0.13\textwidth,trim=0 -10 1 0]{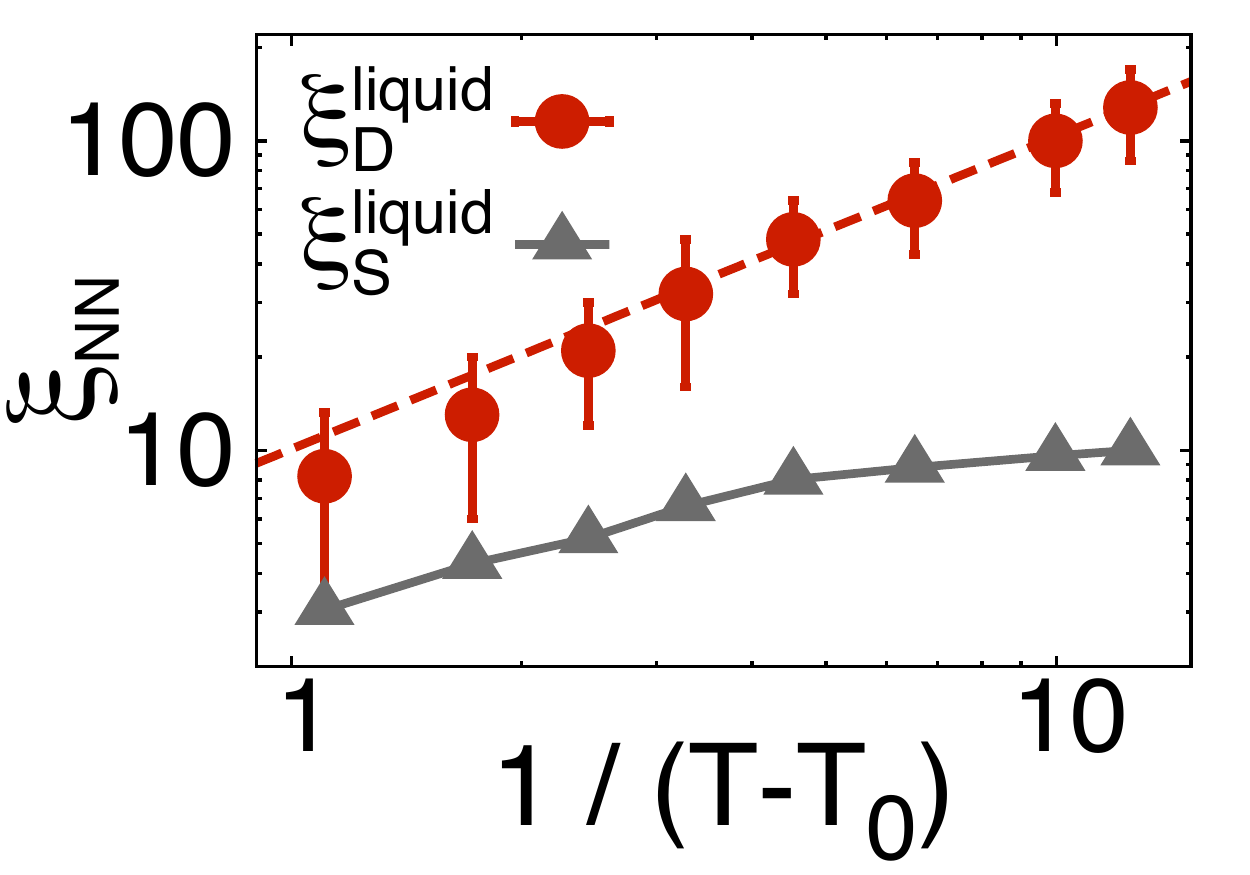}}
\vspace{-.6 cm}
\caption{\label{R8tavg} (color online).
Time-averaged density evolution in the supercooled binary liquid
with $\alpha_i/\beta_i=100$.
$N(\vec{r})$ at $T=1.225$ (top), $0.541$ (middle), 
and $0.420$ (bottom) averaged over the indicated multiples of each
system's $\tau^*_{NN}$.
A sub-cubic section of each cell has been removed to
reveal a portion of the inner simulation.
$n_A(\vec{r})$ time averages are black-orange,
$n_B(\vec{r})$ time averages are black-white.
Bottom Right: Log-plot of static and dynamic correlation lengths 
vs.~$1/(T-T_0)$. The dashed red line has a slope of 1.0.
}
\end{figure}

The effects of this growing dynamic length scale are therefore
especially prominent in highly underdamped systems, where its properties can
be observed and quantified through finite size effects in $F(q,t)$.
For example, when $\alpha_i/\beta_i=100$
finite size effects become numerically insurmountable below $T \simeq 0.4$.
Measurements indicate that the average two point liquid static correlation 
length 
$\xi^{\rm liquid}_S$ grows slowly, approximately as $1/T$, while
the dynamic correlation length grows more rapidly, as
$\xi^{\rm liquid}_D \sim (T-T_0)^{-1\pm0.35}$ (see Fig.~\ref{R8tavg}).
This indicates that the supercooled liquid exhibits
heterogeneous dynamics driven by strong inertial effects.
Similar links between slowing dynamics and growing dynamic correlation
lengths have been widely discussed
\cite{glotzer00,RFOT07,gibbs58,*biroli07,
natureglass10,kimyam00,weeks00,*weeks07}.

A correlation between fragility and the length scale for
cooperativity
is consistent with existing interpretations of
strong and fragile liquids
\cite{angellrev00}.
We also find a relevant link to recent experiments on colloidal glasses
which demonstrate a transition from strong to fragile behavior
as the elastic
properties of the colloidal particles become increasingly stiff
\cite{naturegels09}.
When overdamped, Eq.~(\ref{pfcdyn2}) describes
a very soft, visco-elastic solid, while elastic stiffness and
fragility both increase as damping is reduced.
This is because $\alpha_i \sim v_s \sim \sqrt{E}$, where $v_s$ is
a sound speed and $E$ is the relevant elastic modulus.
Greater elastic stiffness should therefore correspond to reduced effective 
damping and, we expect, increased fragility \cite{Novikov05note}.
This agrees with the trend found in Ref.~\cite{naturegels09}.

Figure \ref{R8tavg} shows simulation images of $N(\vec{r})$ for the 
$\alpha_i/\beta_i=100$ system averaged over various
times at $T=1.225$, $0.541$, and $0.420$.
Caging is apparent at short times for all $T$, but
the long time averages at low $T$ retain more of their original structure
as the peaks exhibit less translational freedom.
It is important to note that
time averages are shown at equal multiples of each liquid's relaxation
time, not at equal $t$, 
so that time scales remain normalized as $T$ is varied.
The continuous but rapid decline in translational freedom as $T$ is lowered
signals a smooth transition from liquid-like to activated dynamics.
This is consistent with the postulated crossover at $T_c$, below which 
relaxations are expected to be limited by increasingly rare, heterogeneously
correlated cage escape events.
This transition coincides with
the emergence of the plateau in $F_{ij}(q,t)$ and the
split second peak in $S_{NN}^P(q)$ below $T \simeq 0.6$,
as shown in Fig.~\ref{dynrange}.

\begin{figure}[btp]
 \centering
 \subfigure{\includegraphics*[width=0.235\textwidth,height=0.175\textwidth]{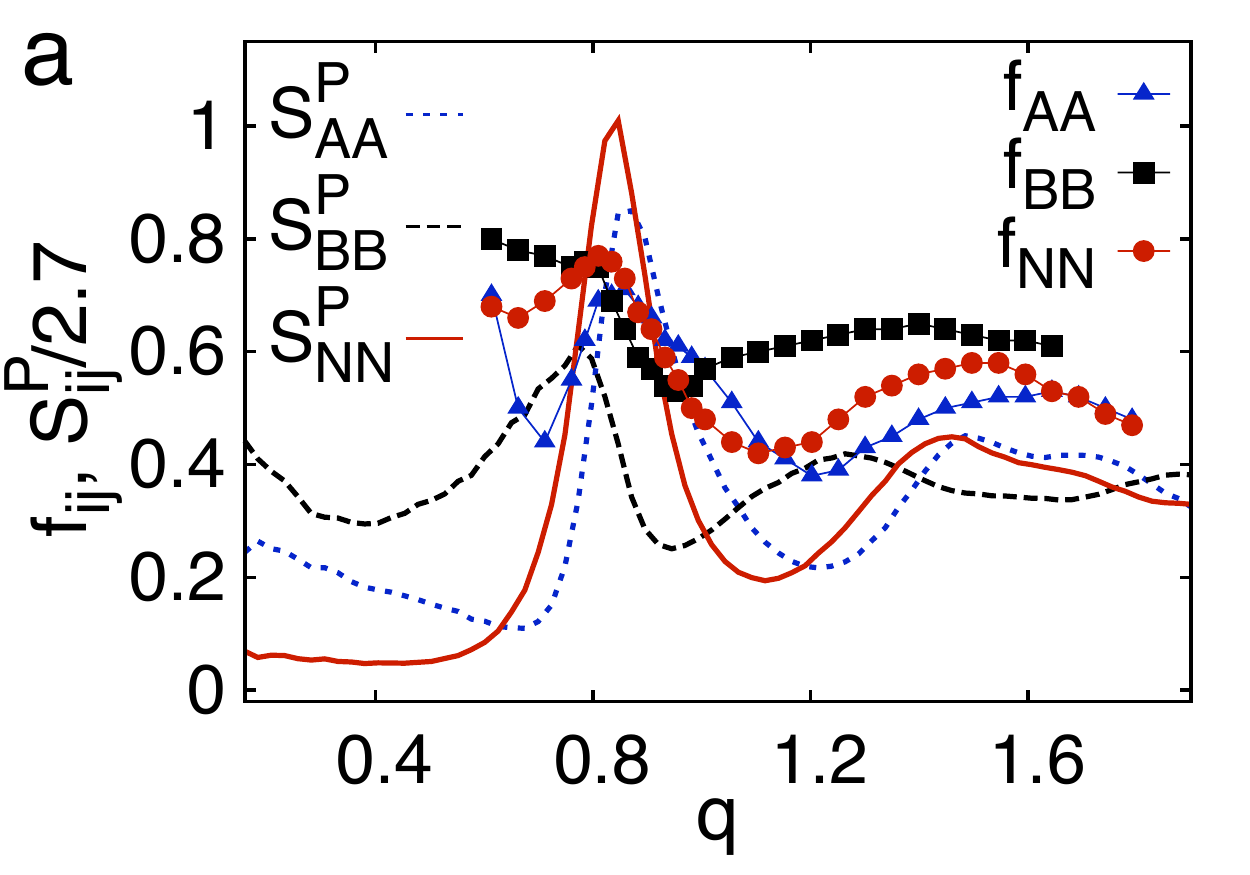}}%
 \hspace{.1 cm}
 \subfigure{\includegraphics*[width=0.235\textwidth,height=0.175\textwidth]{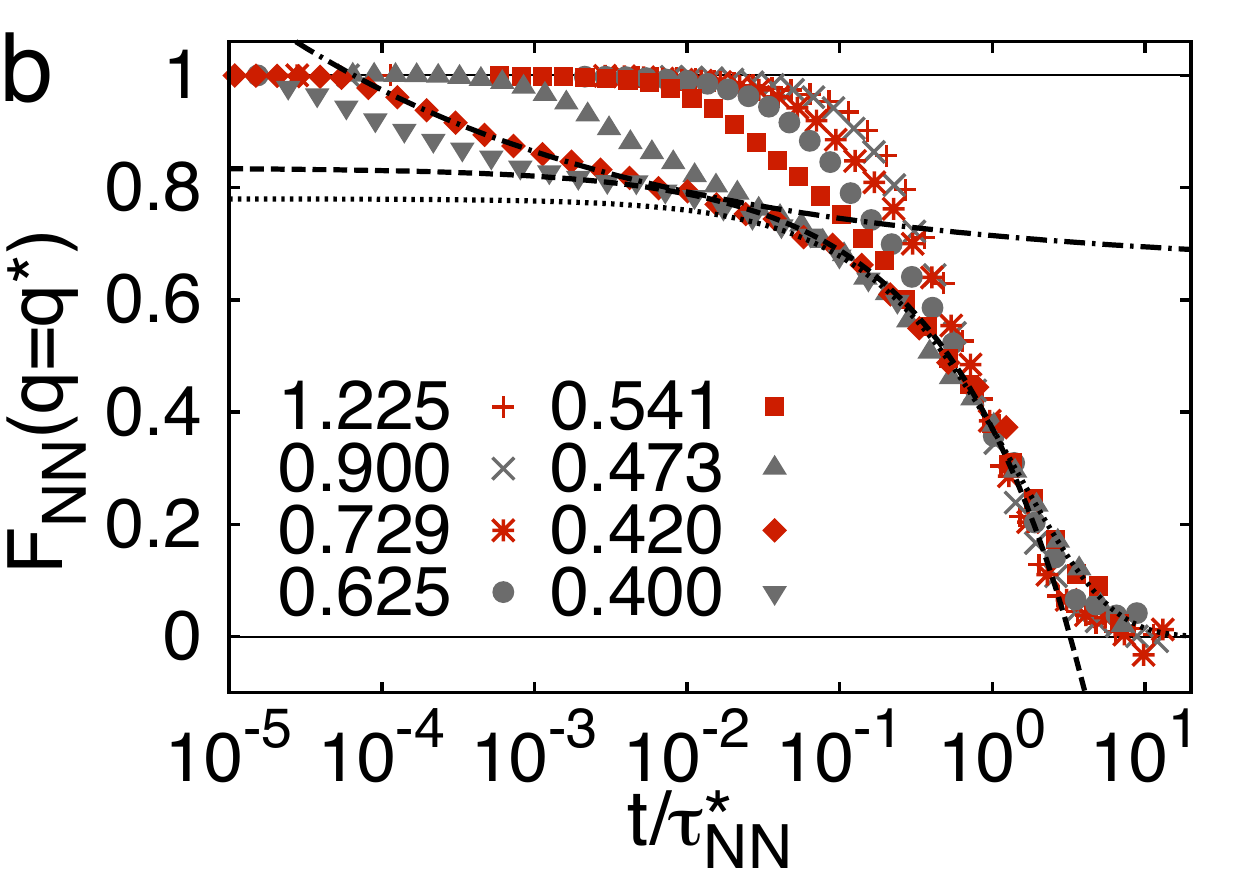}}\\
\vspace{-.3 cm}
\caption{\label{MCTR8} (color online).
The highly underdamped $\alpha_i/\beta_i=100$ model and MCT.
(a) $f_{ij}(q)$ and $S^P_{ij}(q)/2.7$ at $T=0.420$,
(b) $F_{NN}(q,t/\tau^*_{NN})$ with various MCT scaling
functions. Dashed: von Schweidler power law, Dotted: stretched exponential,
Dashed-Dotted: 
critical decay power law.
}
\end{figure}

Figure \ref{MCTR8} shows data for the $\alpha_i/\beta_i=100$ system
relevant for comparison with the predictions of MCT.
The nonergodicity parameter, $f_{ij}(q)$
(height of the plateau in $F_{ij}(q,t)$), 
is plotted in Fig.~\ref{MCTR8}a for $T=0.420$.
It follows the normal MCT behavior in which $f_{ij}$ decays
while oscillating in phase with $S_{ij}(q)$.
Some of the dynamic scaling behaviors predicted by MCT are tested in
Fig.~\ref{MCTR8}b.
Present results indicate that the von Schweidler scaling for late
$\beta$-relaxations (initial decay after plateau),
$F(q,t)=f-B(t/\tau)^b$,
is obeyed reasonably well over 2-3 orders of magnitude in time.
The measured exponent 
$b \simeq 0.45\pm 0.15$ is comparable to typical values.
Fits to the MCT critical decay power law (initial decay to plateau),
$F(q,t)=f+At^{-a}$, give $a \simeq 0.3\pm0.1$.

The late $\alpha$-relaxations predicted by MCT are generally well-approximated
by a stretched exponential decay. Our data are fit quite well by this
form, as shown in Fig.~\ref{MCTR8}b, but with a stretching exponent $\beta$
that decreases with $T$ from approximately $1$ to $0.6$. 
MCT also predicts that the initial divergences
of the fast and slow relaxation times follow power laws,
$\tau_{\beta} \sim (T-T_c)^{-1/(2a)}$
and
$\tau_{\alpha} \sim (T-T_c)^{-\gamma},$
respectively,
where $\gamma=1/(2a)+1/(2b)$.
Fits to these forms are shown in Fig.~\ref{dynrange}e, and though
the Vogel-Fulcher fits are superior, the power laws are reasonably
accurate through the early super-Arrhenius growth.

These results confirm that the DDFT equation of motion
with inertia 
does in fact describe a glass transition, and that when damping is weak 
this transition 
strongly resembles both the structural glass transition observed
for fragile glass formers as well as that predicted by MCT.
Our results are consistent with a picture in which fragility
is driven by a large dynamic correlation length, which
in some cases can be associated with large elastic moduli.
A direct test of this association could be performed using colloidal
systems such as those of Ref.~\cite{naturegels09}. By varying the degree of 
confinement, one could compare the relative magnitudes and
growth rates of any dynamic correlation length as behavior 
is varied from strong to fragile.

\begin{acknowledgments}
The authors acknowledge valuable comments from Ken Elder and Dan Vernon,
and support from the Natural Science and Engineering Research Council of 
Canada, {\it le Fonds Qu\'eb\'ecois de la recherche sur la nature et les 
technologies}, and the Schulich and Carl Reinhardt Endowments.
Simulations were performed on a CLUMEQ/Compute Canada cluster
at the Universit\'e Laval and a Beowulf cluster at McGill University.
\end{acknowledgments}

\end{document}